\documentclass[12pt,showpacs,aps,prd]{revtex4}
\usepackage{graphicx}
\input epsf

\begin{document}
\title{$\eta_{b}BB^{\ast}$ vertex from QCD sum rules}
\author{Chun-Yu Cui$^{\P}$, Yong-Lu Liu$^*$ and Ming-Qiu Huang$^*$}
\affiliation{$^{\P}$ School of Biomedical Engineering, Third Military Medical University and Chongqing University, Chongqing 400038, China}
\affiliation{$^*$ Department of Physics, National University of Defense
Technology, Hunan 410073, China}
\date{\today}
\begin{abstract}
The form factor and the coupling constant in the $\eta_{b}BB^{\ast}$ vertex are evaluated in the framework of three-point QCD sum rules. The correlation functions responsible for the vertex is evaluated considering contributions of both $B$ and $B^{\ast}$ mesons as off-shell states. The form factors obtained are different for the two cases, whereas the final results of the coupling constant are compatible.
\end{abstract}
\pacs{ 11.55.Hx,  13.75.Lb, 13.25.Ft,  13.25.Hw}
\maketitle
%%%%%%%%%%%%%%%%%%%%%%%%%%%%%%%%%%%%%%%%%%%%%%%%%%%%%%%%%%%%%%%%%%%%
%\section{Introduction}
%%%%%%%%%%%%%%%%%%%%%%%%%%%%%%%%%%%%%%%%%%%%%%%%%%%%%%%%%%%%%%%%%%%%

\section{Introduction}
Recently, the bottomonium ground state $\eta_b$ was discovered in the processes $\Upsilon(3S,2S)\to \eta_b+\gamma$~\cite{UP3,UP2}. More data are expected to be accumulated in the forthcoming SuperB and LHC-b.
Therefore, investigation of the spectroscopy and decay processes related to this state becomes rather instructive. Since it is still difficult to study strong interaction phenomena at non-perturbative regime, the study of quarkonium decay is generally performed in the framework of effective Lagrangian with meson exchange. Thus, the reliable determination of various characteristics, such as form factors and coupling constants are needed. In the decay process $\Upsilon(4S,5S)\to \eta_b+\gamma$, rescattering effects are important for understanding the anomalous largeness of the branching ratios~\cite{XiangLiu}. As suggested in their work, this decay proceeds in two steps. First the $\Upsilon(4S,5S)$ decays into a $BB^{\ast}$ intermediate state, and then these two particles produce the final states $\eta_b$ and $\gamma$ by exchanging a $B$ meson. In order to compute the effect of these interactions in the final decay rate, the coupling constant of the $\eta_{b}BB^{\ast}$ vertex is an necessary input parameter. To describe strong interactions of $\eta_{b}BB^{\ast}$ at the hadronic level, the following effective Lagrangian is employed~\cite{XiangLiu}:
\begin{eqnarray}
\mathcal{L}_{\eta_{b}BB^{\ast}}&=& ig_{\eta_{b}BB^{\ast}}B^*_{\mu}\partial^\mu\eta_b{B}^{+}.
\label{1}
\end{eqnarray}
The $\eta_{b}BB^{\ast}$ interactions are characterized by strong coupling constant $g_{\eta_{b}BB^{\ast}}$. However, such low-energy hadron interaction lie in a region which is very far away from the perturbative regime, precluding us to use the perturbative approach with the fundamental QCD Lagrangian. Therefore, we need some non-perturbative approaches, such as QCD sum rules \cite{Shifman,RRY,Nielsen}, to calculate the form factors.

In this article, we calculate the form factor $g_{\eta_{b}BB^{\ast}}(Q^2)$ in the framework of three-point QCD sum rules (QCDSR). More specifically, we evaluate two form factors: one when $B$ is the off-shell particle and another when $B^{\ast}$ is off-shell. The two results are parametrized by analytical forms and then extrapolated to obtain the coupling constant $g_{\eta_{b}BB^{\ast}}$. Herein, we use the same technique for the study of the couplings in the vertices $D^{\ast} D \pi$~\cite{nnbcs00,nnb02}, $D D \rho$~\cite{bclnn01}, $D^{\ast} D^{\ast} \pi$~\cite{cdnn05}, $D^{\ast}D^{\ast}\rho$~\cite{bcnn08}, $D D \omega$~\cite{hmm07}, $D^{\ast}_s D K^{\ast}(892)$~\cite{Azizi10}, $D_s D K^{\ast}_{0}$~\cite{Azizi11} and $B^{\ast}_{s1}B^{\ast}K$~\cite{Cui}.

The outline of the Letter is as follows. In Sec.~\ref{sec2}, we present QCD sum rules for the considered vertex when both $B$ and $B^{\ast}$ mesons are off-shell. Sec.~\ref{sec3} is devoted to the numerical analysis and discussion.

\section{The sum rule for the $\eta_{b}BB^{\ast}$ vertex}\label{sec2}
In this section, we present QCD sum rules for the form factor of the $\eta_{b}BB^{\ast}$ vertex.
The three-point function associated with the $\eta_{b}BB^{\ast}$ vertex,
for an off-shell $B$ meson, is given by
\begin{equation}
\Gamma_{\mu}^{B}(p,p^{\prime})=\int d^4x \, d^4y \;\;
e^{ip^{\prime}\cdot x} \, e^{-iq\cdot y}
\langle 0|T\{j_{\mu}^{B^{\ast}}(x) j^{{B}\dagger}(y)
 j^{{\eta_{b}}\dagger}
(0)|0\rangle,
\label{correboff}
\end{equation}
where the interpolating currents are $j_{\mu}^{B^{\ast}}(x) = \bar q(x) \gamma_{\mu} b(x)$, $j^{B}(x) = i\bar q(x)\gamma_{5}b(x)$, and $j^{{\eta_{b}}}(x)= i\bar b(x)\gamma_{5}b(x)$.
The correlation function for an off-shell $B^{\ast}$ meson is
\begin{equation}
\Gamma_{\mu}^{B^{\ast}}(p,p^{\prime})=\int d^4x \,
d^4y \;\; e^{ip^{\prime}\cdot x} \, e^{-iq\cdot y}\;
\langle 0|T\{j^{B}(x) j_{\mu}^{B^{\ast}\dagger}(y)
 j^{{\eta_{b}}\dagger}
(0)\}|0\rangle,
\label{correroff}
\end{equation}
and $q=p^{\prime}-p$ is transferred momentum.
The general expression for the vertices
(\ref{correboff}) and (\ref{correroff})
can be written in terms of the invariant amplitudes associated with two independent Lorentz structures:
\begin{eqnarray}
\Gamma_{\mu}(p,p^{\prime})&=&
\Gamma_1(p^2 , {p^{\prime}}^2 , q^2) p_{\mu}
+ \Gamma_2(p^2,{p^{\prime}}^2, q^2) p^{\prime}_{\mu}.
\label{trace}
\end{eqnarray}

In compliance with the QCDSR, the above correlation functions need to be calculated in two different ways: in the theoretical side, they are evaluated by the help of the operator product expansion (OPE), where the short and large distance effects are separated; In the phenomenological side, they are calculated in terms of hadronic parameters such as masses, leptonic decay constants and form factors. The sum rules for the form factors are obtained with both representations being matched, invoking the quark-hadron duality and equating the coefficient of a sufficient structure from both sides of the correlation functions. To improve the matching between the two representations, double Borel transformation with respect to the variables, $P^2=-p^2\rightarrow M^2$ and ${P^\prime}^2=-{p^\prime}^2\rightarrow {M^{\prime}}^2$, is performed.

The physical side of correlation function (\ref{correboff}) is studied with hadronic degrees of freedom. From Eq. (\ref{1}), one can deduce the matrix elements associated with the $\eta_{b}BB^{\ast}$ vertex:
\begin{eqnarray}
\langle \bar B(p)\eta_{b}(q)|B^{{\ast}}(p+q,\epsilon)\rangle &=& g_{\eta_{b}BB^{\ast}}
\left(p\cdot\epsilon^{\ast}\right).
\label{ffb}
\end{eqnarray}

The meson decay constants $f_{\eta_{b}}$, $f_{B}$, and $f_{B^{\ast}}$ are
defined by the following matrix elements:
\begin{eqnarray}
\langle0|j^{\eta_{b}}|{\eta_{b}(p)}\rangle&=& \frac{m_{\eta_{b}}^2}{2m_{b}}f_{\eta_{b}},\nonumber\\
\langle 0|j^{B}|{B(p)}\rangle&=& \frac{m_{B}^2}{m_{b}}f_{B},\nonumber\\
\langle 0|j_{\mu}^{B^{\ast}}|{B^{\ast}(p)}\rangle&=& m_{B^{\ast}} f_{B^{\ast}}\epsilon^{\mu}_{B^{\ast}}(p).
\label{fff}
\end{eqnarray}

Saturating Eq.~(\ref{correboff}) by the complete set of appropriate $B^{\ast}$, $B^{\ast}$ and $\eta_{b}$ states, using Eqs.~(\ref{ffb}) and (\ref{fff}), and then summing over polarization vectors via
\begin{eqnarray}
\epsilon^{\mu}_{B^{*}}(p){\epsilon^{\nu}_{B^{*}}}^{*}(p)=-g_{\mu\nu}+\frac{p_{\mu}p_{\nu}}{m_{B^{\ast}}^2},
\label{polvec}
\end{eqnarray}
the physical side of the correlation function for $B$ off-shell is obtained
\begin{eqnarray}
\Gamma_{\mu}^{(B)phen}(p,p^{\prime})&=&\frac{C}
{(P^2+m^2_{\eta_{b}})(Q^2+m^2_{B})({P^\prime}^2 +m^2_{B^{\ast}})}\times\nonumber\\
&&\left[g^{B}_{\eta_{b}BB^{\ast}}(q^2)(-g_{\mu\nu}+\frac{p^{\prime}_{\mu}p^{\prime}_{\nu}}{ m^2_{B^{\ast}}})p_{\nu}\right] \nonumber\\
&&+ ...\,.
\label{phenboff}
\end{eqnarray}

In a similar way, the physical side for an off-shell $B^{\ast}$ meson is obtained as
\begin{eqnarray}
\Gamma_{\mu}^{(B^{\ast})phen}(p,p^{\prime})&=&\frac{C}
{(P^2+m^2_{\eta_{b}})(Q^2+m^2_{B^{\ast}})({P^\prime}^2 +m^2_{B})}\times\nonumber\\
&&\left[g^{B}_{\eta_{b}BB^{\ast}}(q^2)(-g_{\mu\nu}+\frac{q_{\mu}q_{\nu}}{ m^2_{B^{\ast}}})p_{\nu}\right] \nonumber\\
&&+ ...\,,
\label{phenbsoff}
\end{eqnarray}
where we use the abbreviation
\begin{eqnarray}
C&=&\frac{m_{\eta_{b}}^2m_{B}^2m_{B^{\ast}}f_{\eta_{b}}f_{B^{\ast}}f_{B}}
{2m_{b}^2},\nonumber
\end{eqnarray}
and ``..." represents the contribution of the higher states and
continuum.

\begin{figure}
\begin{minipage}{7cm}
\epsfxsize=15cm \centerline{\epsffile{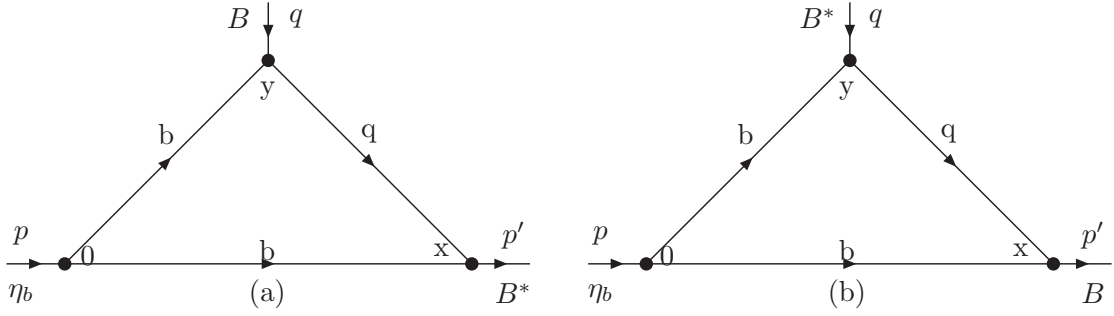}}
\end{minipage}
\caption{{(a) and (b): Bare loop diagrams for the $B$  and
$B^{\ast}$ off-shell, respectively}}\label{Figure1}
\end{figure}

\begin{figure}
\begin{minipage}{7cm}
\epsfxsize=15cm \centerline{\epsffile{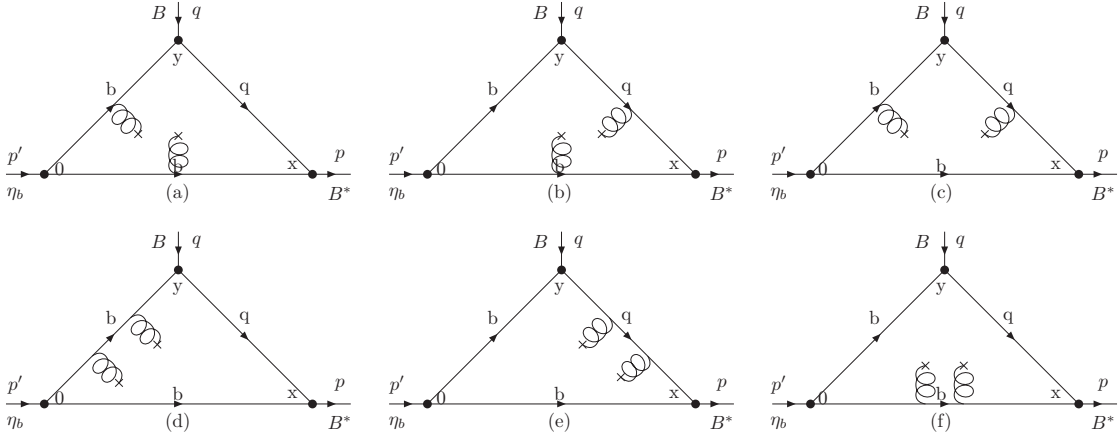}}
\end{minipage}
\caption{{Diagrams for contributions of bi-gluon operator in the case $B$ off-shell.}}\label{Figure2}
\end{figure}

In the following, we turn to the computation of the correlation functions in the QCD side. Each invariant amplitude $\Gamma^{i}(p^{\prime},p)$ appearing in Eqs.~(\ref{trace}) can be written in terms of perturbative and condensate terms
\begin{equation}
\Gamma_{i}=\Gamma_{i}^{per}+\Gamma_{i}^{(3)}+\Gamma_{i}^{(4)}+\Gamma_{i}^{(5)}
+\Gamma_{i}^{(6)}+\cdots
\label{gamma}
\end{equation}
where $\Gamma_{i}^{per}$ is the perturbative contribution, and $\Gamma_{i}^{(3)}$, $\cdots$, $\Gamma_{i}^{(5)}$ are
contributions of condensates of dimension 3, 4, 5, $\cdots$
operators in the OPE.

\begin{figure}
\begin{minipage}{7cm}
\epsfxsize=15cm \centerline{\epsffile{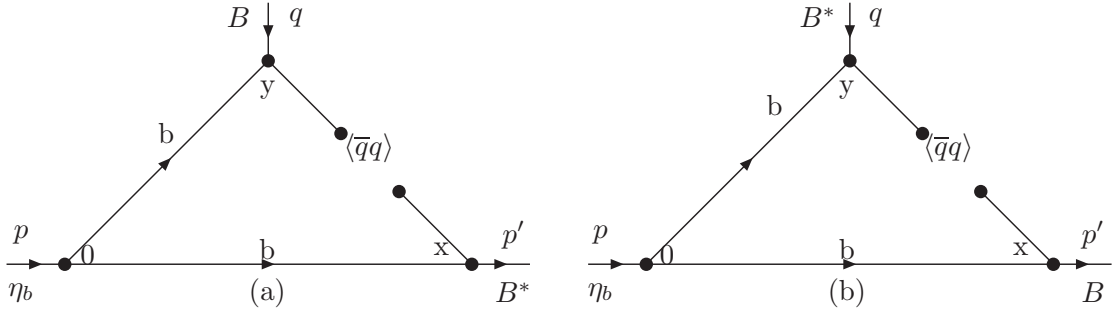}}
\end{minipage}
\caption{{(a) and (b): Diagrams corresponding to quark condensate for the $B$ off-shell and $B^{\ast}$
off-shell, respectively.}}\label{Figure3}
\end{figure}

The perturbative contribution and gluon
condensate contribution can be defined in terms of double dispersion integral as
\begin{eqnarray}
\Gamma_{i}^{per}&=&-\frac{1}{4\pi^2}\int_{s_{min}}^\infty ds
\int_{u_{min}}^\infty du \:\frac{\rho^{per}_i(s,u,Q^2)}{(s-p^2)(u-{p^\prime}^2)}\;,
 \nonumber  \\
\Gamma_{i}^{(4)}&=&-\frac{1}{4\pi^2}\int_{s_{min}}^\infty ds
\int_{u_{min}}^\infty du \:\frac{\rho^{(4)}_{i}(s,u,Q^2)}{(s-p^2)(u-{p^\prime}^2)},
\nonumber  \nonumber\;\;\;\;\;\;i=1,\ldots,14, \label{dis}
\end{eqnarray}
where $\rho^{per}_i(s,u,Q^2)$ is the perturbative spectral density and $t=q^2$. The perturbative spectral density can be obtained by calculating diagrams (a) and (b) in Fig.(\ref{Figure1}). In the calculation, Cutkosky rules are adopted to deal with the usual Feynman integral of these diagrams, i.e., by replacing the quark propagators with Dirac
delta function $\frac{1}{q^2-m^2}\rightarrow (- 2\pi i)
\delta(q^2-m^2)\theta(q^0)$. The physical region in $s$ and $u$ plane is described by the following inequalities:\\
\begin{eqnarray}\label{13au}
-1\leq F^{B}(s,u)=\frac{s^{1/2}((s-u-t)/2+m_{q}^2-m_{b}^2)}
{(s/4-m_{b}^2)^{1/2}\lambda^{1/2}(s,u,t)}\leq+1,\nonumber\\
-1\leq F^{B^{\ast}}(s,u)=\frac{s^{1/2}((s-u-t)/2+m_{q}^2-m_{b}^2)}
{(s/4-m_{b}^2)^{1/2}\lambda^{1/2}(s,u,t)}\leq+1,
\end{eqnarray}
where $\lambda(a,b,c)=a^2+b^2+c^2-2ac-2bc-2ab$ and $t=q^2=-Q^2$.
The diagrams for the contribution of the gluon condensate in the case $B$ off-shell are depicted in Fig.(\ref{Figure2}). We follow the method employed in Refs.~\cite{yangmz,yangmz1}, namely, directly calculate the imaginary part of the integrals in terms of the Cutkosky rules. In this Letter we use the structures $p_{\mu}$ for the off-shell $B$ meson and $p^{\prime}_{\mu}$ for the off-shell $B^{\ast}$ meson. After some straightforward calculations, the spectral densities is obtained as following:
\begin{eqnarray}
\rho_1^{B(per)}(s,t,u)&=&\frac{3}{[\lambda(s,u,t)]^{3/2}}(3*(u(2t(m_{b}^2-m_{b} m_{q}-s)+s(m_{b}-m_{q})^2)\nonumber\\
&&+(m_{b}-m_{q})(s-t)(m_{b}t+m_{q} s)+m_{b}u^2 (m_{q}-m_{b}))),\nonumber\\
\rho_1^{B(4)}(s,t,u)&=&\frac{\langle g^{2}G^{2}\rangle}{2[\lambda(s,u,t)]^{3/2}}(-3s+3t-5u).
\label{SpecDenOSB}
\end{eqnarray}
for the off-shell $B$ meson, and
\begin{eqnarray}
\rho_2^{B^{\ast}(per)}(s,t,u)&=&\frac{3}{[\lambda(s,u,t)]^{3/2}}
s*(m_{b}^2(-(s-t+u))+m_{q}^2(s-t+u)+u(s+t-u)),\nonumber\\
\rho_2^{B^{\ast}(4)}(s,t,u)&=&\frac{\langle g^{2}G^{2}\rangle}{2[\lambda(s,u,t)]^{3/2}}
(s+t-u)
\label{SpecDenOSBs}
\end{eqnarray}
for the off-shell $B^{\ast}$ meson. All powers of the light quark mass are included when calculating the spectral density. In the above expressions, $\rho_{1(2)}$ represents the spectral density for $p_{\mu}(p^{\prime}_{\mu})$.

As what has been shown in~\cite{nnbcs00,Nielsen}, heavy quark condensate contributions are negligible in comparison with the perturbative one. Thus, only light quark condensate contributions to the correlators are calculated with considering the diagrams (a) and (b) of Fig. (\ref{Figure3}). However, contributions of the light quark condensate contributions are zero after the double Borel transformation with respect to the both variables $P^2$ and ${P^\prime}^2$. The $D=5$ quark-gluon mixing condensate contributions are also zero.

Quark-hadron duality assumption is adopted to subtract the contributions of the higher states and continuum, i.e., it is assumed that
\begin{eqnarray}\label{ope}
\rho^{higher states}(s,u) = \rho^{OPE}(s,u,t) \theta(s-s_0)
\theta(u-u_0),
\end{eqnarray}
where $s_0$ and $u_0$ are the continuum thresholds.

The double Borel transformation with respect to $P^2=-p^2\rightarrow M^2$ and ${P^\prime}^2=-{p^\prime}^2\rightarrow {M^{\prime}}^2$ are applied when matching two sides of the correlation function. The final sum rules for the corresponding form factors are obtained as:
\begin{eqnarray}\label{CoupCons-GBBR-Boffshel1}
g^{B(1)}_{\eta_{b}BB^{\ast}}(Q^2)&=&\frac{(Q^2+m_{B}^2)}{C}
\mbox{e}^{\frac{m_{\eta_{b}}^2}{M^2}}\mbox{e}^{\frac{m_{B^{\ast}}^2}{{M^{\prime}}^2}}
\left[\frac{1}{4~\pi^2}\int^{s_0}_{4m_{b}^2}
ds\int^{u_{0}}_{m_{b}^2} du\right.\nonumber\\
&&
\left.(\rho_1^{B(per)}(s,t,u)+\rho_1^{B(4)}(s,t,u))\theta[1-F^{B}(s,u)^2]
e^{\frac{-s}{M^2}}e^{\frac{-u}
{{M^{\prime}}^2}}\right]
\end{eqnarray}
and
\begin{eqnarray}\label{CoupCons-GBBR-Bstaroffshel2}
g^{B^{\ast}(2)}_{\eta_{b}BB^{\ast}}(Q^2)&=&\frac{2m_{B^{\ast}}^2(Q^2+m_{B^{\ast}}^2)}{(m_{B}^2-m_{\eta_{b}}^2-t)C}
\mbox{e}^{\frac{m_{\eta_{b}}^2}{M^2}}\mbox{e}^{\frac{m_{B}^2}{{M^{\prime}}^2}}
\left[\frac{1}{4~\pi^2}\int^{s_0}_{4m_{b}^2}
ds\int^{u_{0}}_{m_{b}^2} du\right.\nonumber\\
&&
\left.
(\rho_2^{B^{\ast}(per)}(s,t,u)+\rho_2^{B^{\ast}(4)}(s,t,u))\theta[1-F^{B^{\ast}}(s,u)^2]
e^{\frac{-s}{M^2}}e^{\frac{-u}
{{M^{\prime}}^2}}\right].
\end{eqnarray}

It is noted that in the following analysis, we use the relations between the Borel masses $M^2$ and ${M^{\prime}}^2$ as $\frac{M^2}{{M^{\prime}}^2} = \frac{m^2_{\eta_{b}}}{m^2_{B^{\ast}}}$ for a  $B$ off-shell and $\frac{M^2}{M'^2} = \frac{m^2_{\eta_{b}}}{m^2_{B}}$ for a $B^{\ast}$ off-shell.

\section{Numerical analysis}\label{sec3}
In the numerical analysis of the sum rules, input parameters are shown in Table~\ref{table1}. We take $m_{\eta_{b}}$ and $f_{\eta_{b}}$ from Ref.~\cite{Rashed}. The continuum thresholds, $s_0$ and $u_0$, are not completely arbitrary as they are correlated to the
energy of the first excited states with the same quantum numbers as the states we concern. They are given by $s_0=(m_{\eta_{b}} + \Delta_{s})^2$ and $u_0=(m+\Delta_{u})^2$, where $m$ is the $B^{\ast}$ meson mass for the case that $B$ is off-shell and the $B$ meson mass for that $B^{\ast}$ is off-shell.  $ \Delta_u $ and $ \Delta_s$ are usually around $0.5 \; \mbox{GeV}$. The threshold $s_{0}$, $u_{0}$ and Borel variable $M^{2}$ are varied to find the optimal stability window where OPE convergence, stability and pole dominance of the sum rule with the Borel mass parameter are satisfied.
\begin{table}[h]
\caption{Parameters used in the calculation.}
\begin{center}
\begin{tabular}{ccccccccc}
\hline
$m_{b} (\mbox{GeV})$ &  $m_{B} (\mbox{GeV})$ &$m_{B^{\ast}} (\mbox{GeV})$ & $m_{\eta_{b}} (\mbox{GeV})$ &  $f_{B}(\mbox{GeV})$~\cite{Jamin} &  $f_{B^{\ast}}(\mbox{GeV})$~\cite{Belyaev}  & $f_{\eta_{b}} (\mbox{GeV})$ \\ \hline
$4.2 \pm 0.1$&5.28 &5.325 &9.4 &$0.21 \pm 0.01$ &$0.16 \pm 0.01$ &$0.705 \pm 0.027$ \\ \hline
\end{tabular}
\end{center}\label{table1}
\end{table}

\begin{figure}
% Requires \usepackage{graphicx}
\begin{center}
\begin{minipage}[c]{0.5\textwidth}
\centering
\includegraphics[width=\textwidth]{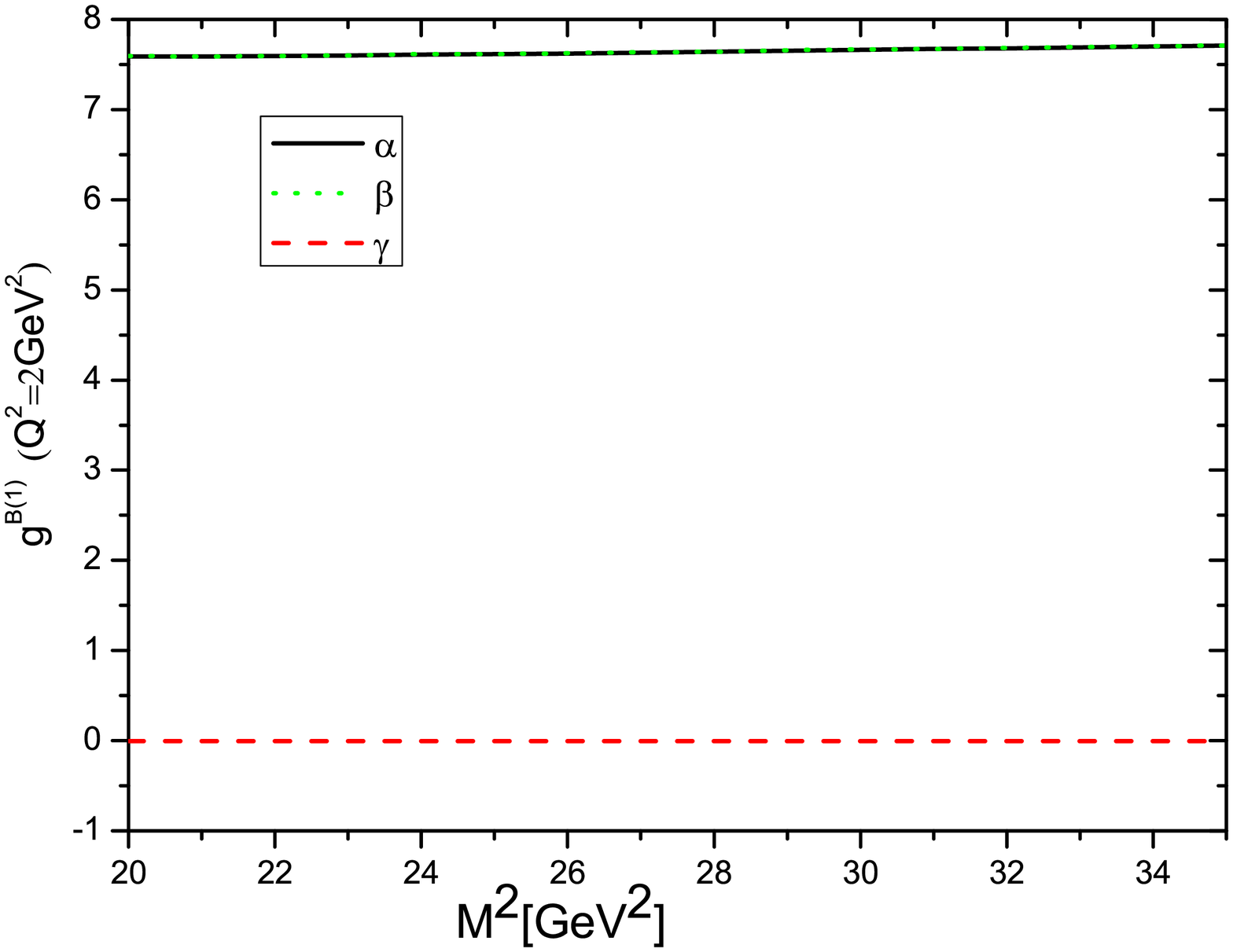}\\
(a)
\end{minipage}
\hspace{-0.1\textwidth}
\begin{minipage}[c]{0.5\textwidth}
\centering
\includegraphics[width=\textwidth]{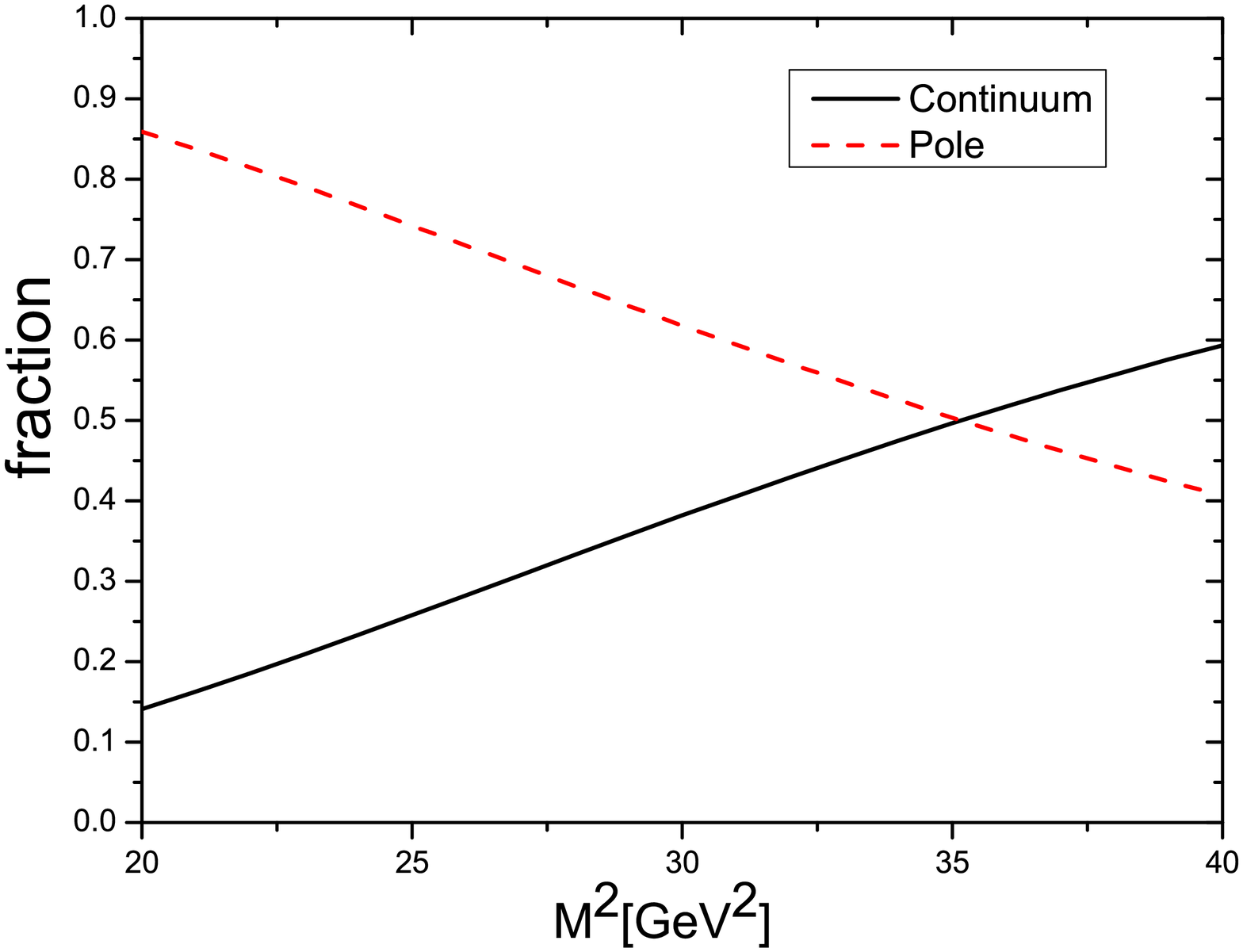}\\
(b)
\end{minipage}
\end{center}
\caption{a) The dependence of the form factor $g^{B(1)}_{\eta_{b}BB^{\ast}}(Q^2=2.0\,\mbox{GeV}^2)$ on Borel mass $M^2$ for $\Delta_{s} =0.5\,\mbox{GeV}$ and $\Delta_{u} =0.5\,\mbox{GeV}$. The notations $\alpha$, $\beta$ and $\gamma$ correspond to
total, perturbative and four-quark condensate contributions respectively and b) pole-continuum contributions.}\label{Figure4}
\end{figure}

Using $\Delta_{s}=\Delta_{u} = 0.5\,\mbox{GeV} $ for the continuum thresholds and fixing $Q^2=2\,\mbox{GeV}^2$, the dependence of the $g^{B(1)}_{\eta_{b}BB^{\ast}}(Q^2=2\,\mbox{GeV}^2)$ on Borel mass are shown in Fig.~(\ref{Figure4}a). From this figure, we see the results exhibit a good OPE convergence and stability for $M^2\geq 20\,\mbox{GeV}^2$. Fig.~(\ref{Figure4}b) demonstrates the contribution from the pole term with variation of the Borel mass $M^2$. We see that the pole contribution is larger than continuum one for $M^2\leq 35\,\mbox{GeV}^2$. We choose $M^2= 33\,\mbox{GeV}^2$ as a reference point.

\begin{figure}
% Requires \usepackage{graphicx}
\includegraphics[width=15cm]{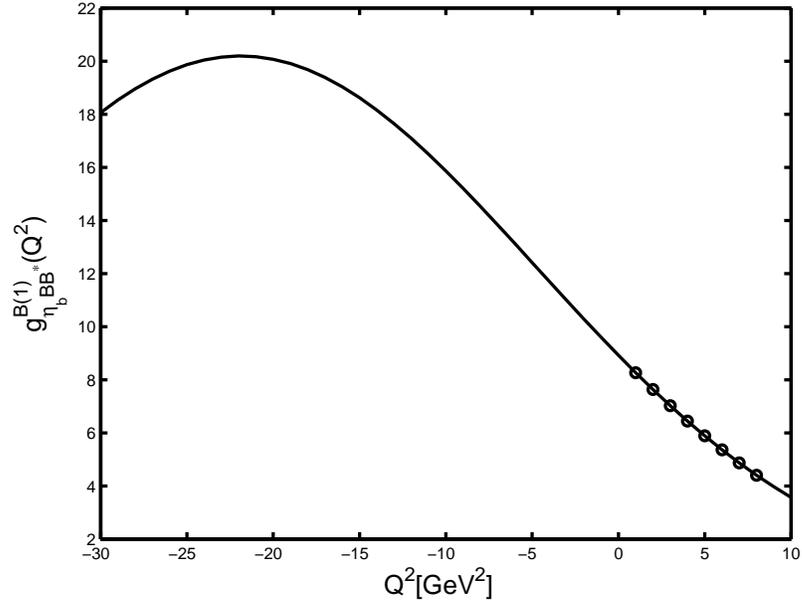}
\caption{$g^{B(1)}_{\eta_{b}BB^{\ast}}(Q^2)$ (circles) QCDSR form factors as a function of $Q^2$. The solid lines
correspond to the Gaussian parametrizations.}\label{Figure5}
\end{figure}

Now, we would like to discuss the behavior of the form factors in terms of $Q^2$, which are plotted in Fig.~\ref{Figure5}. In Fig.~\ref{Figure5}, the circles correspond to the form factor $g^{B(1)}_{\eta_{b}BB^{\ast}}(Q^2)$ in the interval where the sum rule is valid. For $g^{B(1)}_{\eta_{b}BB^{\ast}}(Q^2)$, our result is better extrapolated by the Gaussian fit parametrization,
\begin{eqnarray}\label{gboff}
g^{B(1)}_{\eta_{b}BB^{\ast}}(Q^2)=20.2~
\,\mbox{Exp}\Big(\frac{-(Q^2+21.9\,\mbox{GeV}^2)^2}{587.5\,\mbox{GeV}^4}\Big).
\end{eqnarray}
The coupling constant is defined as the value of the form factor at $Q^2=-m^2$, where $m$ is the mass of the off-shell meson. Using $Q^2=-m_{B}^2$ in Eq.(\ref{gboff}), the coupling constant is obtained as $g^{B(1)}_{\eta_{b}BB^{\ast}}=19.0$.

\begin{figure}
% Requires \usepackage{graphicx}
\begin{center}
\begin{minipage}[c]{0.5\textwidth}
\centering
\includegraphics[width=\textwidth]{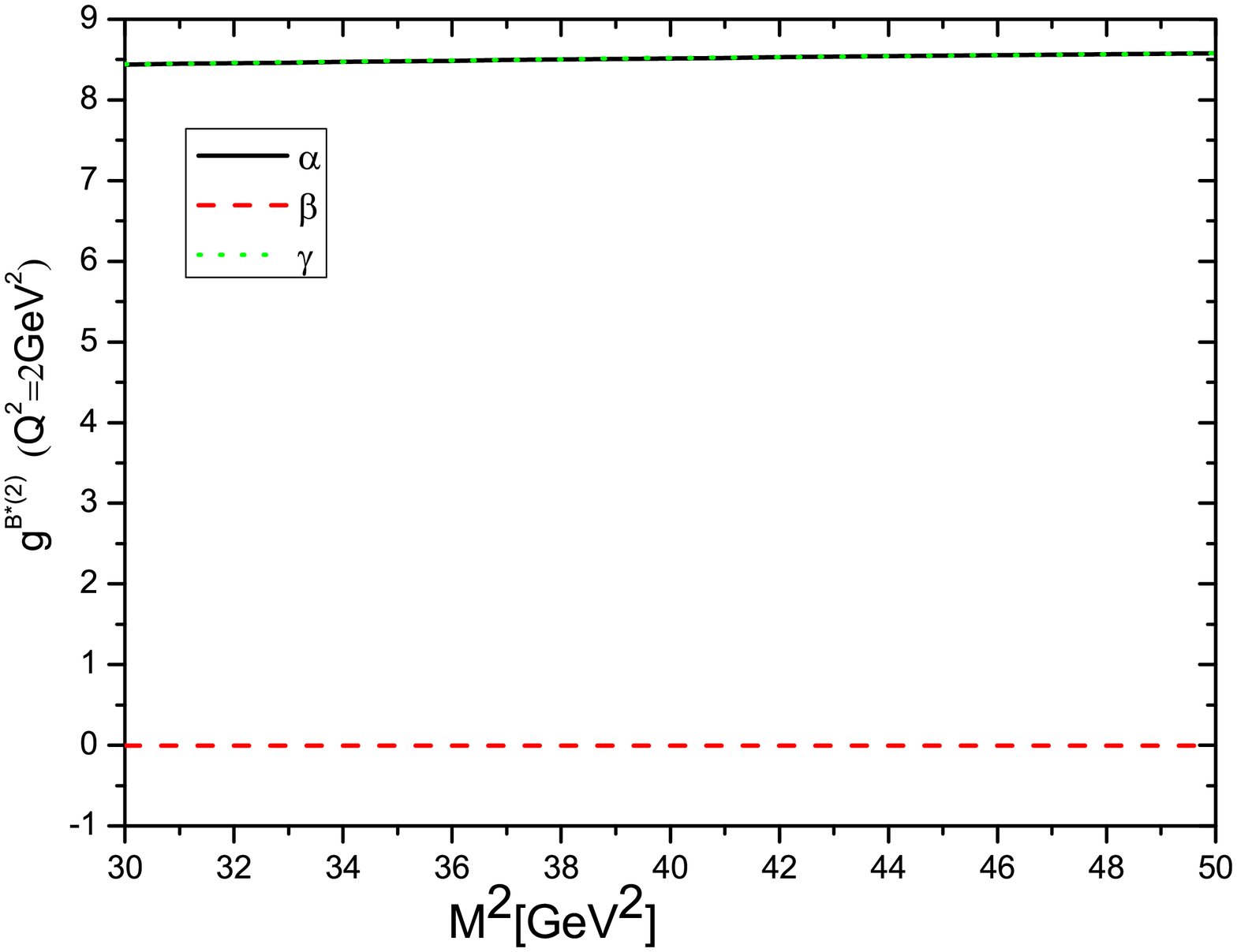}\\
(a)
\end{minipage}
\hspace{-0.1\textwidth}
\begin{minipage}[c]{0.5\textwidth}
\centering
\includegraphics[width=\textwidth]{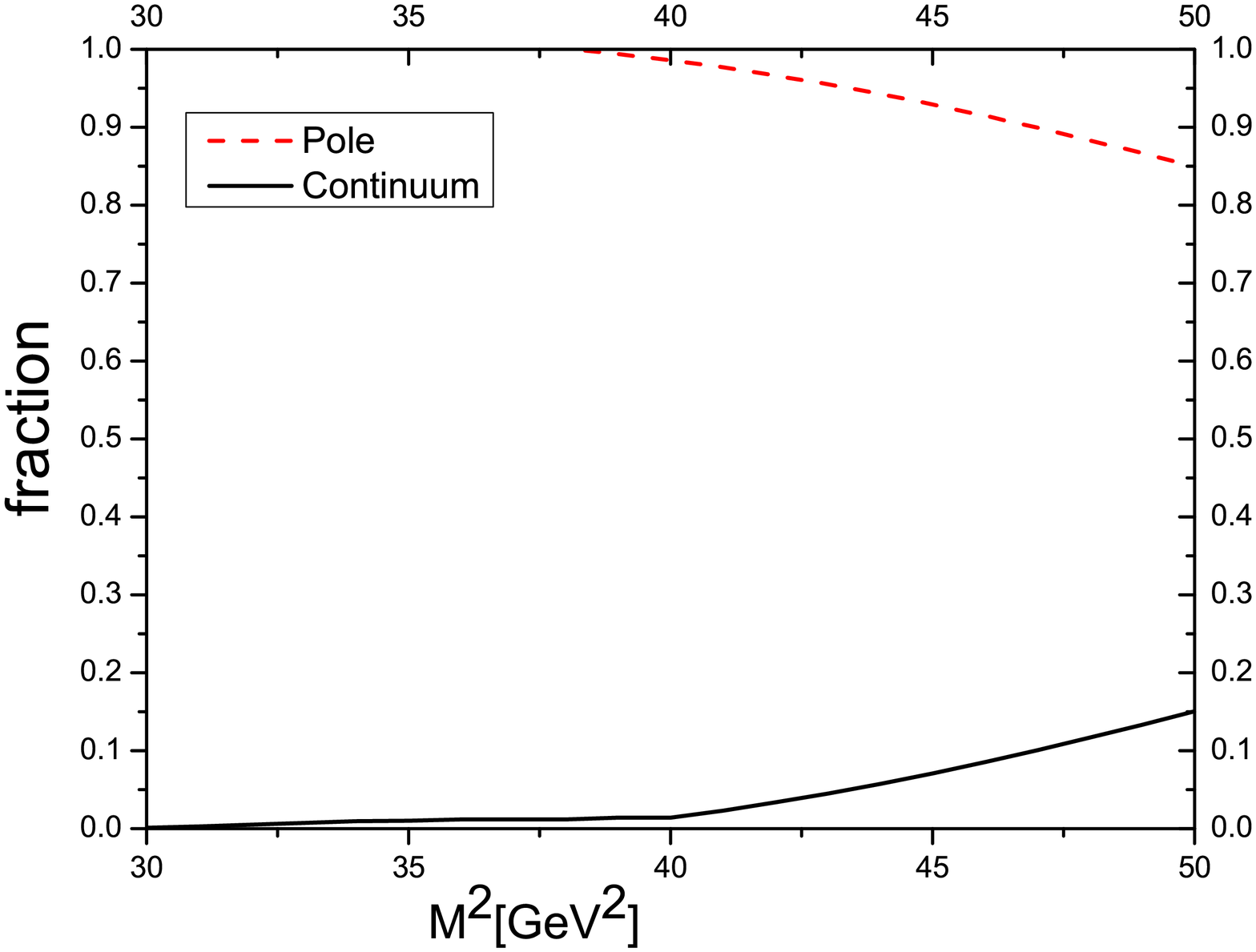}\\
(b)
\end{minipage}
\end{center}
\caption{a) The dependence of the form factor $g^{B^{\ast}(2)}_{\eta_{b}BB^{\ast}}(Q^2=2.0\,\mbox{GeV}^2)$ on Borel mass parameters $M^2$ for $\Delta_{s} =0.5\,\mbox{GeV}$ and $\Delta_{u} =0.5\,\mbox{GeV}$. The notations $\alpha$, $\beta$ and $\gamma$ correspond to
total, perturbative and four-quark condensate contributions respectively and b) pole-continuum contributions.}\label{Figure6}
\end{figure}

\begin{figure}
% Requires \usepackage{graphicx}
\includegraphics[width=15cm]{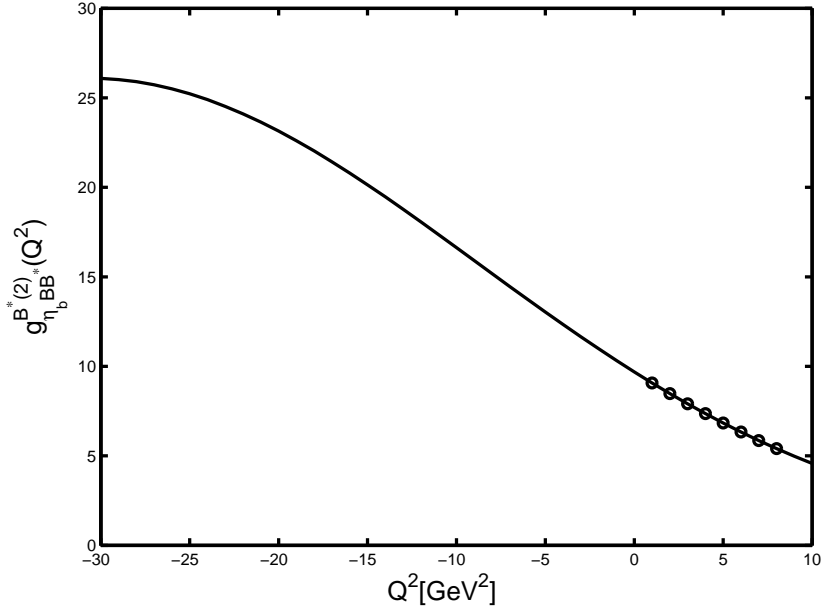}
\caption{$g^{B^{\ast}(2)}_{\eta_{b}BB^{\ast}}(Q^2)$ (circles) QCDSR form factors as a function of $Q^2$. The solid lines correspond to the Gaussian parametrizations.}\label{Figure7}
\end{figure}

In the case that $B^{\ast}$ is off-shell, Fig.~(\ref{Figure6}a) demonstrates a good OPE convergence and stability of $g^{B^{\ast}(2)}_{\eta_{b}BB^{\ast}}$ with respect to the variation of Borel mass for $M^2\geq 30\,\mbox{GeV}^2$. We see that the pole contribution is bigger than the continuum one when the Borel mass $M^2\leq 50\,\mbox{GeV}^2$ from Fig.~(\ref{Figure6}b). Fixing $M^2= 40\,\mbox{GeV}^2$, our numerical results can be fitted by the Gaussian fit parametrization
\begin{eqnarray}\label{gbsoff}
g^{B^{\ast}(2)}_{\eta_{b}BB^{\ast}}(Q^2)=26.1~
\,\mbox{Exp}\Big(\frac{-(Q^2+30.7\,\mbox{GeV}^2)^2}{951.6\,\mbox{GeV}^4}\Big).
\end{eqnarray}
shown by the solid line in Fig.(\ref{Figure7}). The coupling constant $g^{B^{\ast}}_{\eta_{b}BB^{\ast}}=25.8$ is obtained at $Q^2=-m_{B^{\ast}}^2$ in Eq. (\ref{gbsoff}). The two parametrization processes lead to the compatible coupling constant, which is a check on the extrapolation procedure. Taking the average of the two results, we get
\begin{eqnarray}\label{CoupConstg}
g_{\eta_{b}BB^{\ast}}=22.4\pm3.4.
\end{eqnarray}

Following the procedure of error estimate in Refs.~\cite{Nielsen1,Nielsen2}, with all parameters kept fixed, except one which is changed according to its intrinsic error, we calculate a new coupling constant and its deviation. Then we obtain percentage deviation related with each parameter and how sensitive this value is with respect to each parameter. Table~\ref{table2} show the percentage deviation for the two cases. Taking into account the uncertainties presented in the tables, we finally obtain
\begin{eqnarray}
g_{\eta_{b}BB^{\ast}}=24.0\pm10.2.
\end{eqnarray}

\begin{table}[h!]
\begin{center}
\begin{tabular}{|c|c|c|c|}\hline
               &    \multicolumn{2}{|c|}  {\textbf{Deviation {\%}}} \\ \hline
\textbf{Parameters} &\textbf{ $B$ off-shell}& \textbf{$B^{\ast}$ off-shell} \\ \hline
  $f_{B}=210 \pm 10 $ (MeV)   &   $3.9$    & 15.2     \\\hline
  $f_{B^*}=160 \pm 10 $ (MeV) &   $11.6$    & 15.2     \\\hline
  $f_{\eta_{b}}=705 \pm 27 $ (MeV)  & $21$ & $7.3 $    \\\hline
  $m_b=4.20\pm 0.1$ (GeV) & $19.1$ & $ 27.8$           \\\hline
  $M^2 \pm 10\%$ (GeV)& $4.7$ & $1.7$                  \\\hline
  $\Delta s \pm 0.1$ e $\Delta u \pm 0.1 $(GeV)  & $38.9$ &$ 23.5$  \\\hline
\end{tabular}
\caption{Percentage deviation related with each parameter for $g_{\eta_{b}BB^{\ast}}$. }
\label{table2}
\end{center}
\end{table}

We noticed that, in Refs.~\cite{XiangLiu,Meng:2007tk,Meng:2008bq}, the value $g_{\eta_{b}BB^{\ast}}=g_{\Upsilon(1S) BB}=15$ is used as an input parameter. Their estimate is based on the assumption of the heavy quark spin symmetry. Considering the uncertainties, our results are compatible with their estimates. However, our calculation shows that the predicted central value from QCDSR is $60\%$ larger than their result. This may be a demonstration of some degree of heavy quark spin symmetry violation. The numerical value of this parameter may affect the physical observable to some extent. Thus, we expect a further theoretical investigation may give a better confirmation.

In conclusion, we have used three-point QCDSR to calculate the form factor of the $\eta_{b}BB^{\ast}$ vertex. Both cases that $B$ is off-shell and $B^{\ast}$ is off-shell have been considered. As a side product of the form factor, the coupling constant is estimated.
%%%%%%%%%%%%%%%%%%%%%%%%%%%%%%%%%%%%%%%%%%%%%%%%%%%%%%%%%
\section*{Acknowledgement}
This work was supported in part by the National Natural Science
Foundation of China under Contract Nos.11275268 and 11105222.
%%%%%%%%%%%%%%%%%%%%%%%%%%%%%%%%%%%%%%%%%%%%%%%%%%%%%%%%


\begin{thebibliography}{99}
\bibitem{UP3} B. Aubert {\it et al.} Phys. Rev. Lett. {\bf 101}, 071801 (2008).
\bibitem{UP2} B. Aubert {\it et al.} Phys. Rev. Lett. {\bf 103}, 161801 (2009).
\bibitem{XiangLiu} Hong-Wei Ke, Xue-Qian Li and Xiang Liu, Phys. Rev. D {\bf 82}, 054030 (2010).
\bibitem{Shifman}  M. A. Shifman, A. I. Vainshtein and V. I. Zakharov, Nucl. Phys. B 147, 385 (1979).
\bibitem{RRY} L.J. Reinders, H. Rubinstein and S. Yazaki, Phys. Rept.
{\bf 127}, 1 (1985).
\bibitem{Nielsen} M.~E.~Bracco, M.~Chiapparini, F.~S.~Navarra, M.~Nielsen, [arXiv:hep-ph/1104.2864v1]
\bibitem{nnbcs00} F.S. Navarra {\it et al.},
Phys. Lett.  B {\bf 489}, 319 (2000).
\bibitem{nnb02} F. S. Navarra, M. Nielsen, M. E. Bracco, Phys. Rev. D {\bf 65}, 037502 (2002).
\bibitem{bclnn01} M. E. Bracco {\it et al.}
Phys. Lett. B {\bf 521}, 1 (2001).
\bibitem{cdnn05}  F.~Carvalho, F.~O.~Dur\~aes, F.~S.~Navarra and M.~Nielsen,
                  Phys. Rev. C {\bf 72}, 024902 (2005).
\bibitem{bcnn08} M.~E.~Bracco, M.~Chiapparini, F.~S.~Navarra, M.~Nielsen, Phys. Lett. B {\bf 659}, 559 (2008).
\bibitem{hmm07}  L.~B.~Holanda, R.~S.~Marques de Carvalho and A.~Mihara,
                 Phys. Lett. B {\bf 644}, 232 (2007).
\bibitem{Azizi10} K. Azizi and H. Sundu, J. Phys. G: Nucl. Part. Phys. {\bf 38}, 045005 (2011).
\bibitem{Azizi11} H. Sundu, J.Y. Sugu, S. Sahin, N. Yinelek and K. Azizi, Phys. Rev. D {\bf 83}, 114009 (2011).
\bibitem{Cui} Chun-Yu Cui, Yong-Lu Liu and Ming-Qiu Huang, Phys. Lett. B {\bf 707} 129 (2012); Phys. Lett. B {\bf 711} 317 (2012).
\bibitem{yangmz}D.~S.~Du, J.~W.~Li and M.~Z.~Yang, Eur. Phys. J. C {\bf 37},  173 (2004)
\bibitem{yangmz1}M.~Z.~Yang, Phys. Rev. D {\bf 73}, 034027 (2006)
\bibitem{Rashed} Ahmed Rashed, Murugeswaran Duraisamy and Alakabha Datta,
                 Phys. Rev. D {\bf 82}, 054031 (2010).
\bibitem{Jamin} Matthias Jamin and Bjorn O. Lange, Phys. Rev. D {\bf 65}, 056005 (2002).
\bibitem{Belyaev} V. M. Belyaev, V. M. Braun, A. Khodjamirian and R. Ruckl, Phys. Rev. D {\bf 82}, 054031 (2010).

\bibitem{Nielsen1} B.~Os\'orio Rodrigues, M.~E.~Bracco, M.~Nielsen and F.~S.~Navarra, Nucl. Phys. A {\bf 852}, 127 (2011).
\bibitem{Nielsen2} A.~Cerqueira Jr., B.~Os\'orio Rodrigues and M.~E.~Bracco, [arXiv:hep-ph/1109.2236v1]
\bibitem{Meng:2007tk} C.~Meng and K.~T.~Chao, Phys. Rev. D {\bf 77}, 074003 (2008)

\bibitem{Meng:2008bq} C.~Meng and K.~T.~Chao, Phys. Rev. D {\bf 78}, 074001 (2008)

\end{thebibliography}
\end{document}